# Forced and Self Rotation of Magnetic Nanorods Assembly at the Cell Membrane: a Bio-Magnetic Torsion Pendulum

*François Mazuel, Samuel Mathieu, Riccardo Di Corato, Jean-Claude Bacri, Thierry Meylheuc, Teresa Pellegrino, Myriam Reffay[*], Claire Wilhelm[*]*

Dr. F. Mazuel, S. Mathieu, Dr. R. Di Corato, Prof. J.-C. Bacri, Dr. M. Reffay, Prof. C. Wilhelm
Laboratoire Matière et Systèmes Complexes (MSC), UMR 7057, CNRS and Université Paris Diderot, 75205 Paris Cedex 05, France.
E-mails: myriam.reffay@univ-paris-diderot.fr, claire.wilhelm@univ-paris-diderot.fr

Dr. R. Di Corato
Dipartimento di Matematica e Fisica "Ennio De Giorgi", Università del Salento
Via Arnesano, 73100 Lecce, Italy

Dr. T Meylheuc
INRA, UMR1319 Micalis, F-78352 Jouy-en-Josas, France

Prof. T. Pellegrino
Istituto Italiano di Tecnologia, I-16163 Genoa, Italy

**Abstract**

In order to give insights into how anisotropic nano-objects interact with living cell membranes, and possibly self-assemble, we designed magnetic nanorods with average size around 100 nm x 1μm by assembling iron oxide nanocubes within a polymeric matrix under a magnetic field. We then explored the nano-bio interface at the cell membrane under the influence of a rotating magnetic field. We observed a complex structuration of the nanorods intertwined with the membranes. Unexpectedly, after a magnetic rotating stimulation, the resulting macrorods were able to rotate freely for multiple rotations, revealing the creation of a bio-magnetic torsion pendulum.

In addition to their recognized potential for advancing diagnostics and therapies, nanomagnetic materials were recently described as tools to probe or act on the cell membrane. In magnetic manipulation of cell membranes, an external rotating magnetic field is used to act at a distance, with precisely controlled intensity, direction and localization of the applied magnetic torque. Magnetically driven rotating or oscillating magnetic nanoparticles could thus be used to probe cells' mechanical properties,[1-3] to deliver drugs,[4-8] or to kill cells by physical membrane rupture.[9-16] In this last application, anisotropic nanoparticles in disk- or rod-like shapes, stimulated with low-frequency magnetic fields, could compromise the cell membrane and thereby trigger apoptotic or necrotic programmed cell death. Another promising and recent field of research takes advantage of field-mediated magnetic nanoparticle bioassembly to activate biochemical signaling mechanisms.[17-23] When localized to the cell membrane,[17-21] the resulting clustering or orientation of targeted receptors can be seen as a nanomagnetic switch to trigger cell responses.

In parallel, chemical synthesis has yielded magnetic nanomaterials with anisotropic geometries, often rod-like shapes. [15, 24] Synthesis or self-assembly of magnetic particles under a magnetic field is then frequently used to form super-organized anisotropic structures. [25-29] However, how magnetic anisotropic nano-objects interact and assemble with cell membranes under the action of a remote magnetic field has rarely been investigated. Here, we addressed this field by producing highly magnetic nanorods, by assembling them at the cell membrane into larger macrorods through remote spinning by a rotating magnetic field, and by exploring the macrorod bio-structuration and dynamic response.

*Magnetic macrorod formation at the cell membrane*

Nanorods were prepared by self-aggregation of iron oxide nanocubes (20-nm edge) embedded within a polymeric matrix (loaded with FITC dye) in the presence of a 0.05 T magnetic field.

The resulting fluorescent magnetic nanorods (**Figure 1A1**) are (1.1±0.1) μm long and (110±15) nm wide (Figure 1A1) and exhibit strong magnetization (80 emu/g). The nanorods spontaneously attach to the cell membrane (**Figure 1A2**) and are then internalized when incubated overnight in absence of a magnetic field (**Figure 1A3**). However, if incubation takes place in a magnetic field (**Figures 1B** and **Figure 1C**), internalization is prevented. The nanorods are then forced to assemble at the membrane into larger anisotropic structures aligned in the direction of the magnetic field, which we call macrorods. The formation of such macrorods is a two-step process. First, dipolar magnetic interactions (created by a 0.2-T static vertical magnetic field) forces the assembly of multiple nanorods into thin micron-long elongated clusters (Figure 1B) during their first stage of attachment to the cell membrane. In the second step, cells are subjected to a spinning 0.28-T magnetic field rotating around a 45°C cone, for 30 min at 1 Hz. This conical rotating field is created by two permanent magnets that generate a horizontal 0.2-T magnetic field in-between the magnets and spin around the sample, itself within the vertical 0.2-T magnetic field (see experimental. The thin nanorod clusters then merge, forming larger ellipsoid "macrorods" (average length: 22±11μm and diameter: 4±1μm). SEM and confocal images of typical macrorods are shown in Figures 1C1-3.

First step was to evaluate the impact of the magnetic stimulations on cell viability (Figure 1D). It demonstrated that neither the static magnetic field nor the rotating one result in any cell damage, as quantified by measuring the cells metabolic activity (relative to untreated control cells) the day after the magnetic stimulations. The nanorods / cell membrane interactions thus do not induce any significant alteration of cells viability and proliferation capacity, even in the rotating setting, contrary to some other studies reporting membrane physical rupture triggered by rotating magnetic nanoparticles, generally targeted to a specific receptor.[9-16] Here it shows that cells can also adapt to a rotating stress and avoid massive harm.

*Free rotation*

Right after the rotating stimulation, we removed the cells from the magnetic fields (rotating and static) and observed them under the microscope. As expected from the cell viability measurements, cells morphologies were similar to control, with no cell rounding or detachment. However, and surprisingly, some of the macrorods continued rotating, with no external energy source (magnetic or other) to explain this motion (**Figure 1D**, see also supplementary **Movie S1** – accelerated 5 times). This "spontaneous" rotation reveals in fact the remarkable adaptability of cell membrane materials. Indeed, the macrorods rotated in the direction opposite to the magnetic field rotation to which the rods were exposed. This phenomenon is impressive, as it implies that the cell membranes confer a very efficient elastic energy storage ability on the macrorods. This energy accumulated during the forced magnetic excitation is release by free rotations in the opposite direction when the stimulation is stopped (hundreds of free rotations were sometimes recorded).

*Macrorods: intertwined nanorods and membrane filaments*

The explanation for this elastic response comes first from the microscopic structure of the macrorods. Macrorods are composed not only of nanorods but are mingled with membrane structures all along their axis. This is clearly illustrated by the typical confocal image of a macrorod in **Figure 2A**, with the cell membranes labeled with the red Pkh26 membrane marker, and nanorod fluorescence collected in the green channel (FITC). Note that the membrane labels are present all along the macrorod. Looking at the same samples with scanning electron microscopy (SEM, **Figure 2B**) an intertwining of membrane filaments and nanorods are observed (Figure 2B2). Membrane fragments are also detected on transmission electron microscopy (TEM) images (**Figure 2C**) of macrorods detached from cells.

The use of the rotating magnetic conical field is decisive to structure the anchor point of the nanorods within the membrane. It is responsible for membrane-rods entanglement and for radial organization of membrane filaments around the macrorod. Indeed, these hybrid membrane/nanorod structures are absent when rods are assembled without rotation (Figure 2B1).

*Biomagnetic macrorods at the cell membrane: a torsion pendulum?*

To model and quantify the observed free rotation, the rotating stimulation process was explored. A programmable electromagnetic rotating field was used to impose the same rotating cone under the microscope as that generating macrorod formation. Both the magnetic rotating stimulation and the free relaxation could then be video-monitored (supplementary **Movie S2** – accelerated 5 times). Because the field intensity was now lower (30 mT) and thus not sufficient to form the macrorods, they were first formed as previously described with rotating permanent magnets (130 mT). The electromagnetic field was programmed to spin the macrorods counterclockwise, for a given number of rotations (generally 20), at 2 Hz. When the forcing was stopped, and thus in absence of a magnetic field, the macrorods rotated clockwise, some of them up to 20 times, while others rotated only a few degrees (**Figure 3**). We propose a torsion pendulum model to describe this free rotation.

The membrane structure at the anchor point during the forced rotation generates an elastic torque C and the rod is subjected to a rotational viscous friction $\xi$ when it moves. As inertia can be neglected (low Reynolds number $\approx 10^{-2}$), the rod's angle $\theta$ obeys the equation $\xi \frac{d\theta}{dt} = -C(\theta - \theta_R)$, with $\theta_R$ the total of relaxation rotation (in degrees), and can thus be simply written as $\theta(t) = \theta_R \left(1 - \exp\left(-\frac{t}{\tau}\right)\right)$ with $\tau = \frac{\xi}{C}$. The fitting in Figure 3A demonstrates the correct matching of this torsion pendulum model. The parameters $\tau$ and $\theta_R$ reflect the peculiar coiling of cellular membranes around and within the macrorod. Figure 3B shows the plot of $\tau$ as a function of $\theta_R$ for different rods. Two groups emerge: the first one corresponds to rods relaxing only few tens of degrees (black symbols), while the second corresponds to rods relaxing through more than a thousand degrees (> 3 rotations). For this second group, the rod's anchor point is organized such that it enables extremely efficient energy storage during forced rotations as the membrane filament tension increases. In contrast, for the first group, either the organization is inappropriate or the coiling structure may be damaged during the forced rotation, resulting in an abrupt energy release that would prevent free rotation.

Interestingly, the typical relaxation curve shown in Figure 3A presents a succession of angular jumps that follow the exponential trend described above. This could be explained by different coiling modes of the membrane at the anchor point. Jumps would be associated with twisted membrane filaments, whereas the overall relaxation would correspond to membrane filaments coiled around the rod. The supplementary movie S2 illustrate these two modes.

Within the framework of the torsion pendulum model, successive excitations of a rod is expected to damage the anchor point organization as tension increases and exceeds the membrane filament resistance. This is illustrated by Figure 3C. In this case, rods were subjected to successive periods of magnetic field rotation and immobilization. The intensity of the magnetic field in the immobilized phase is half the intensity of the rotating field, so that a rod with a proper coiling at the anchor point can relax if the elastic restoring force is stronger than the magnetic torque. In the particular example shown in Figure 3C, this is the case for the two first excitations: the rod is able to relax for one rotation despite the magnetic field (indicated with black stars). The increase in tension against the magnetic forcing at the anchor point is materialized in Figure 3C with events (indicated with red crosses) when the rod slows down and stops before being caught up by the magnetic field after an additional rotation. The occurrence of such events increases with successive excitations until the structure at the anchor is damaged (red arrow). After that the rod follows perfectly the magnetic field: the damaged structure cannot store anymore elastic energy efficiently.

Taken together, all those observations confirm the simple torsion pendulum model based on a specific membrane coiling that enable energy storage and restitution.

In summary, by remote assembly of magnetic nanorods during their first stage of attachment to cell membranes through the application of a rotating magnetic field, we managed to create a super-assembly (macrorod) trapping within and wrapping around membrane filaments. As a result, this biomagnetic structure exhibited an elastic behavior which provided an impressive numbers of free rotations, when the rotating field is released and no other stimulation applied. This movement, evidences a remarkable pool of membranes available and the possibility to arrange them in a biomagnetic torsion pendulum.

**Experimental section**

*Magnetic nanorods preparation*

The preparation method is based on iron oxide nanocubes assembly into a polymeric matrix, in presence of a static magnetic field, as adapted by a procedure previously described for spherical magnetic nanoparticles [29]. Briefly, the polymeric anisotropic construct is obtained by mixing nanocubes with the poly(maleic anhydride alt-1 octadecene) polymer, in chloroform. The nanoparticles dispersion was placed in an ultrasound bath under the influence of two opposite permanent magnets placed against each other on the vial wall, generating a 0.05 T magnetic field. The slow and controlled addition of acetonitrile induces a change in the solubility of both the nanocubes and polymer and promotes their aggregation. The application of a permanent magnetic field promoted the formation of elongated superstructure. The magnetic nanorods were finally magnetically sorted and resuspended in water.

The iron oxide nanocubes, about 20 nm in cube-edge, were synthesized by thermal decomposition technique detailed as detailed in [30]. Briefly, they were prepared by mixing 1 mmol of iron(III) acetylacetonate and 4 mmol of decanoic acid in 25 mL of dibenzyl ether, heating the solution to 200°C (5°C/min) for 2.5 h. The temperature is then increased further to reflux temperature (300°C at a rate of 10°C/min) for 1 h. At the end of the process, nanocubes are dispersed in chloroform.

*Magnetic field stimulations*

Different magnetic devices were designed and fabricated. The simplest one dedicated to macrorods formation (**Figure 4A**) consists in a permanent magnet (neodymium, 50x40x20 mm, Supermagnet) placed below the cell sample. It creates a magnetic field of 130 mT in the sample region. The second one (**Figure 4B**) is adapted to this first one by adding a set of two permanent magnets facing one to the other and fixed to a motor controlling a gear wheel (frequency up to 2 Hz). This set of magnets creates a horizontal field of 130 mT, so that the resulting field (about 180 mT) makes an angle of 45° with the vertical axis. As a result, when the motorized magnets are rotated, the magnetic field spins, describing a 45° cone. Both devices contain a cylindrical 20 mm cradle to welcome the cell sample, and are thermostated by water circulation. The third device (**Figure 4C**) was designed to be adapted to a microscope and to allow switching on and off the magnetic field at will. It is composed of four coils which cores are made of soft iron, connected by pairs and supplied by an alternative current. The space between each core is about 1cm and it creates in between a 30 mT magnetic field. To generate a rotating 45°C rotating cone, this magnetic device is placed 2 cm beneath the cell sample. The magnetic field on the cells is then reduced to 10 mT. To generate the rotating field in the plane of the magnetic coils, the two pairs of coils are supplied with sinusoidal currents (2A amplitude) displaying the same frequency (up to 5 Hz) but 90° out of phase. Finally, the whole set-up was mounted on the vertical arm of a Leica DMIRB with control of the z position, and the microscope was thermostated at 37°C by cube&box (Life Imaging Services).

*Nanorods incubation with cells and macrorods formation*

PC3 cancer cells were cultured in DMEM supplemented with 10% fetal bovine serum, L-glutamine and penicillin/streptomycin. Prior to experiments, cells were transferred into 20 mm cylindrical home-made cuves with glass bottom which fit within the magnetic fields devices. When cells reached 80-90 % of confluency, nanorods were incubated at an iron concentration of 1 mM, corresponding roughly to $10^6$ nanorods per ml, or equivalently to $10^3$ nanorods per cell. This concentration was adjusted by testing a range of concentration, observing the cells with confocal microscopy, and selecting concentration where nanorods were numerous on the cells membrane, but still quasi-individual.

After the 2-hour incubation period, the cells were immediately transferred into the thermostated magnetic field devices, first the permanent static field for 30 min, then the spinning magnetic field for 30 min.

Cell viability was assessed by Alamar Blue metabolic assay (Thermo Fisher Scientific). Fluorescence (appearing post-metabolization of the active ingredient, Resaruzin), was quantified with a microplate reader (excitation 550 nm, detection 590 nm). In brief, 100 000 cells were first seeded in the cuves (3 per conditions). 24 hours after, nanorods incubation was performed (2-hours at $1mM_{Fe}$, except for the control conditions) and the cuves were submitted either to the static magnetic field only (30 min, condition B in Figure 1), or to both the static and rotating magnetic field (30 min + 30 min, condition C in Figure 1). 24 hours after, all cuves were incubated (800 μl total) with 10 % Alamar Blue in DMEM for 2 hours, and the reagent was transferred to 96-well plate for analysis (200 μl per well). All values are expressed relative to control (normalized at 100% viability).

*Imaging*

Nanorods and macrorods were imaged by scanning (SEM) and transmission (TEM) electron microscopy, by confocal microscopy, and by conventional transmission microscopy. TEM was used to observe nanorods in aqueous suspension, or macrorods extracted from the cells. For this second case, the cells were first fixed with paraformaldehyde (2%) for 30 min right after the whole stimulation protocol. Culture medium was then washed and replaced with ultra-pure water to avoid salts contaminations, and cells were scratched in order to detach cells and macrorods. For both nanorods and macrorods TEM observation, a small drop of suspension (5µl) was pipetted onto a copper grid, and observed after total evaporation of the liquid with a Phillips Tecnai 12.

SEM was used to observe whole cells. Cells were fixed with glutaraldehyde (2% in 0.1 M cacodylate buffer), and dehydrated by soaking in a graded series of ethanol before critical-point drying under $CO_2$. Samples were mounted on aluminum stubs with conductive silver paint and sputter coated with gold palladium for 200 s at 10 mA. Samples were then imaged with a Hitachi S4500 instrument.

For confocal microscopy, cells were labeled either by Pkh26 label (20 min incubation, according to manufacturer's instruction, Sigma) which binds to the plasma membrane, or with phalloidin to see actin filaments, or with DAPI to image the nuclei. Cells were observed by means of an Olympus JX81/ BX61 Device/Yokogawa CSU Device spinning disk microscope (Andor Technology plc, Belfast, Northern Ireland), equipped with a 60x Plan-ApoN oil objective lens.

Conventional transmission microscopy was carried out with a DMIRB Leica microscope, equipped with a Cube & Box device to maintain the cells at 37°C during experiment. Macrorods imposed and free rotations were captured with an ultra-fast camera. Time sampling was used in between 100 to 10 images per second. Image J home-made plugins (succession of thresholding, selecting and measuring) were used to analyze the macrorods rotational movements.

**Acknowledgements**

This work was supported by the European Union (ERC-2014-CoG project MaTissE 648779).

We acknowledge the ImagoSeine core facility of the Institut Jacques Monod, member of IBiSA and France-BioImaging (ANR-10-INBS-04) infrastructures.

Received: ((will be filled in by the editorial staff))
Revised: ((will be filled in by the editorial staff))
Published online: ((will be filled in by the editorial staff))

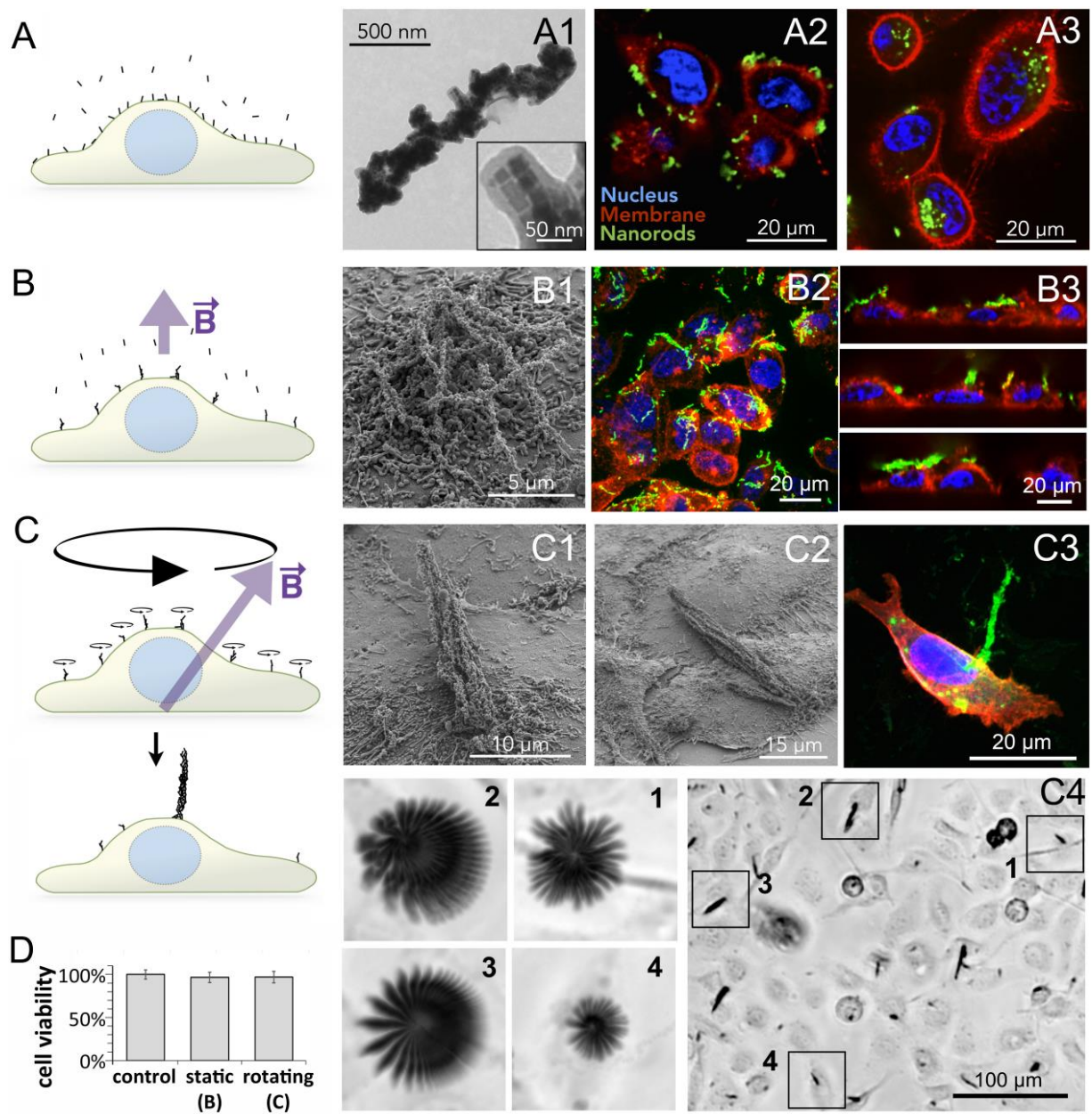

**Figure 1 (2 columns / 16 cm):** Assembly process of magnetic nanorods into a macrorod at the cell membrane. **A**. Nanorods dispersed in the cells culture medium interact individually with the cell membrane without magnetic field. (A1) shows transmission electron microscopy (TEM) image of the magnetic nanorods before cellular interaction (in aqueous dispersion) and identifies iron oxide nanocubes embedded in polymer (insert). (A2) shows confocal image after 1 hour incubation, where nanorods (FITC, green channel) are detected on the cell

membranes (Pkh26, red channel). Nuclei are labeled with DAPI (blue channel). (A3) shows confocal image of the nanorods after a 24h incubation and illustrates their complete internalization in cell cytoplasm. **B**. Formation of small clusters still attached to the membrane under the application of a static vertical magnetic field. (B1) shows a scanning electron microscopy (SEM) picture of these nanorods clusters on cell membrane; (B2-3) shows confocal microscopy of nanorods clusters spread on all membranes (nanorods in green, membranes in red, nuclei in blue): 4-µm width stacked image (B2) or Z views (B3: reconstruction from Z stacks acquired with a 0.5 µm interslice with Image J Volume Viewer plugin, with 45° x tilt angle). **C**. Rotating stimulation process: the magnetic field spins, describing a cone around its initial vertical direction. This rotating motion forces the nanorods clusters to interact over wider range, and form a larger magnetic macrorod. (C1) and (C2) show SEM images of macrorods attached to cell membranes. (C3) shows confocal microscopy (3D stack reconstruction) of a macrorod (green) fixed by one end on the cell surface (here F-actin in red – phalloidin staining). C4 illustrates the free rotations of the macrorods: large view (right) of the cells at the end of the magnetic field rotation process, and superimposition of 60 images (separated by 1s) for the 4 zones delimited by a square (left). It clearly shows that these macrorods rotate, freely, in absence of magnetic field stimulation. The corresponding movie can be seen as supplementary movie S1. **D**. Cell viability measured by quantifying the metabolic activity (Alamar blue) of cells incubated with the magnetic nanorods in presence of the static magnetic field (condition "static", similar to part B), and under the influence of the rotating magnetic field (condition "rotating", similar to part C), and compared to control cells (seeded at the same exact number of cells, see Experimental Section).

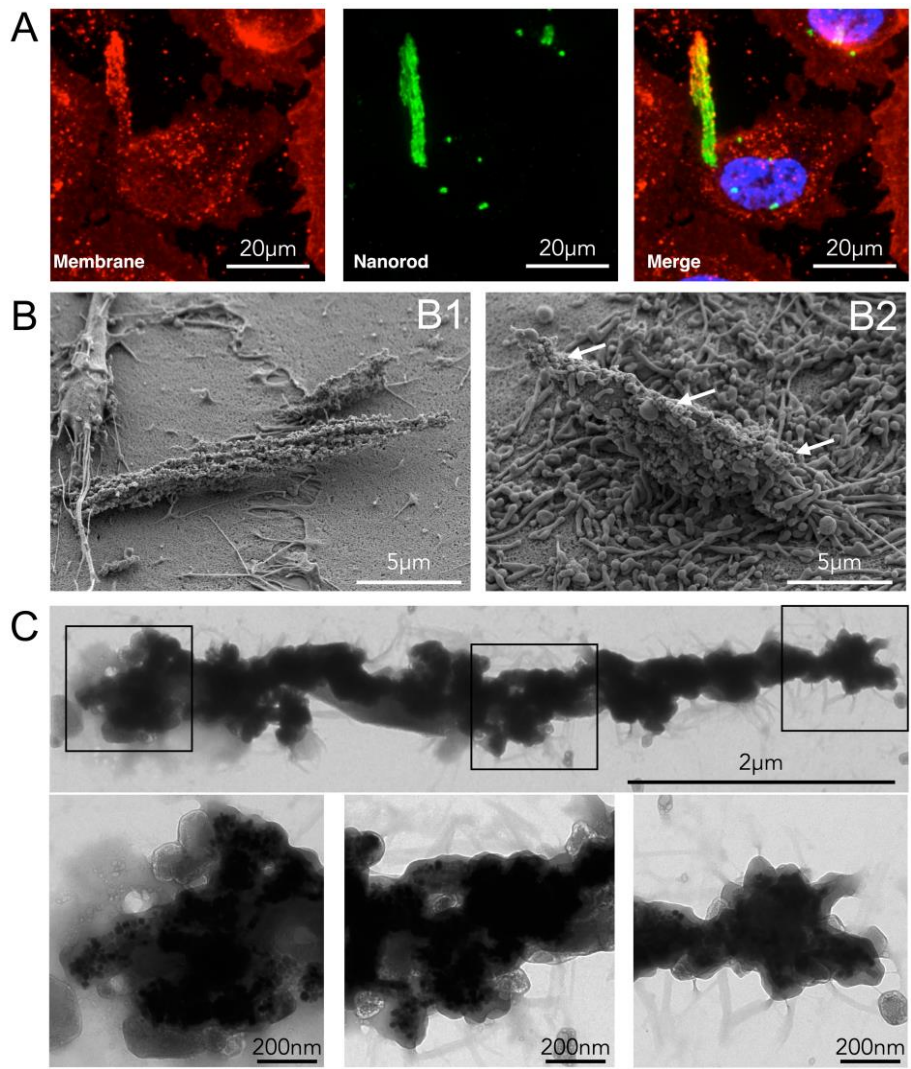

**Figure 2 (1.5 column / 12 cm)**: **A**. Confocal microscopy image of a macrorod in the red channel (membrane, left), green channel (nanorods, middle), and superposition of both (plus nucleus in the blue channel, right). One can clearly see that membranes are trapped all through the macrorod. **B**. SEM pictures of two macrorods either formed with magnetic field rotation (B2), or without (B1). On the right image, membrane filaments are present all along the macrorod (arrows). By contrast, on the left image, it corresponds to "clean" nanorods clusters without membrane entanglements. **C**. TEM pictures of a magnetic macrorod after removal from cell surface. Membrane fragments are detected all around the rod, stuck between the nanorods. This macrorod was probably fixed to the cell by its left end.

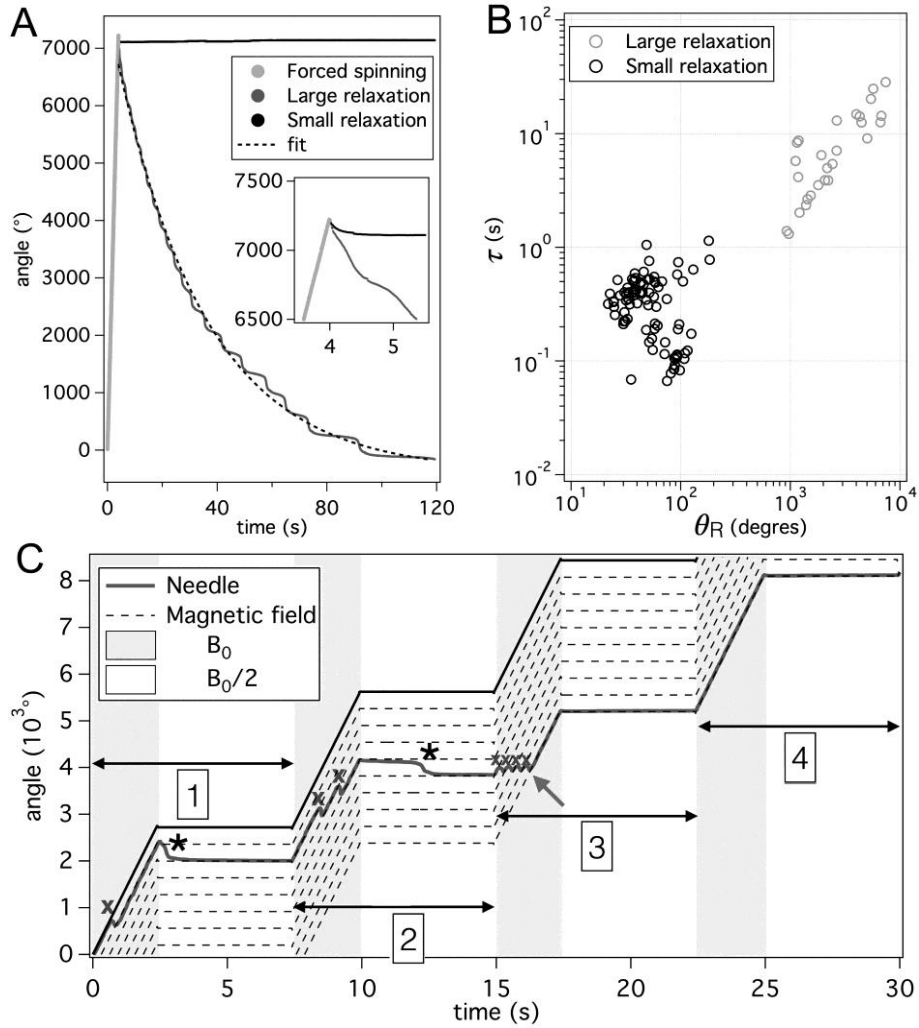

**Figure 3 (1.5 column / 12 cm)**: **A**. Typical angular evolution of two macrorods after the same forced excitation (20 rotations at 2Hz). One relaxes on 20 rotations (grey) whereas the other relaxes only a few degrees (black). The inset shows a magnification around the time when the magnetic stimulation stops. Both relaxation curves are fitted with $\theta(t) = \theta_R \left(1 - \exp\left(-\frac{t-t_0}{\tau}\right)\right)$. The fitting curve is superimposed (dotted line). **B**. Fitting parameters $\tau$ as a function of $\theta_R$ for all analyzed macrorods. Two populations of macrorods emerge. **C**. Successive coiling of the anchor point and its damage. The spinning excitation follows the periodic pattern: 8 counterclockwise rotations at 3.2Hz followed by 5 seconds where the magnetic field position is fixed. Four stimulation cycles are applied. The cumulative angle of a typical macrorod (grey) as well as the magnetic field position modulo

360° (black) are plotted as a function of time. Of note, the intensity of the spinning magnetic field is twice the intensity of the fixed magnetic field. Stars (★) indicate the macrorod relaxations. Instants when the macrorod does not follow the magnetic field are marked with crosses (x). The arrow indicates the time when the anchor point is probably damaged. After that the macrorod follows perfectly the rotating magnetic field.

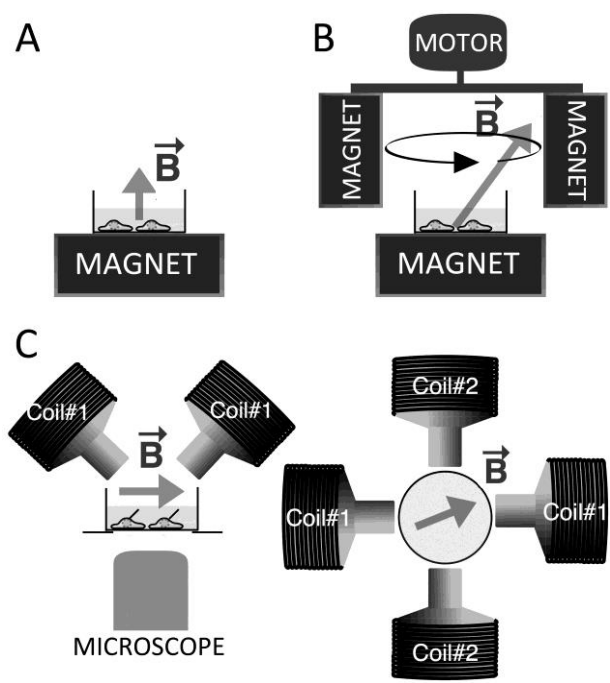

**Figure 4 (one column / 8 cm)**: Magnetic set-ups. **A**. For macrorods formation: a home-made culture dish is placed on a permanent magnet generating a 30 mT vertical magnetic field for 30 min. **B**. The dish is then placed in between two magnets fixed on a motorized axis generating the spinning conical magnetic field (130 mT). **C**. Electromagnetic set-up designed to manipulate the macrorods once they are formed: two pairs of coils are arranged perpendicularly to create a rotating magnetic field in an upper plane over the dish, resulting in a spinning field in the plane of the dish (left = side view / right = top view).

**Table of contents entry**

Assembling nanorods at the cell membrane by a remote spinning magnetic field leads to a complex bio-magnetic macrorod constituted of both nanorods and intertwined cell membrane. This new object behaves as a torsion pendulum able to store elastic energy: it is able to freely rotate over multiple turns in response to the spinning stimulation.

**Keywords:** magnetic nanorods, assembling, membranes, rotating magnetic field

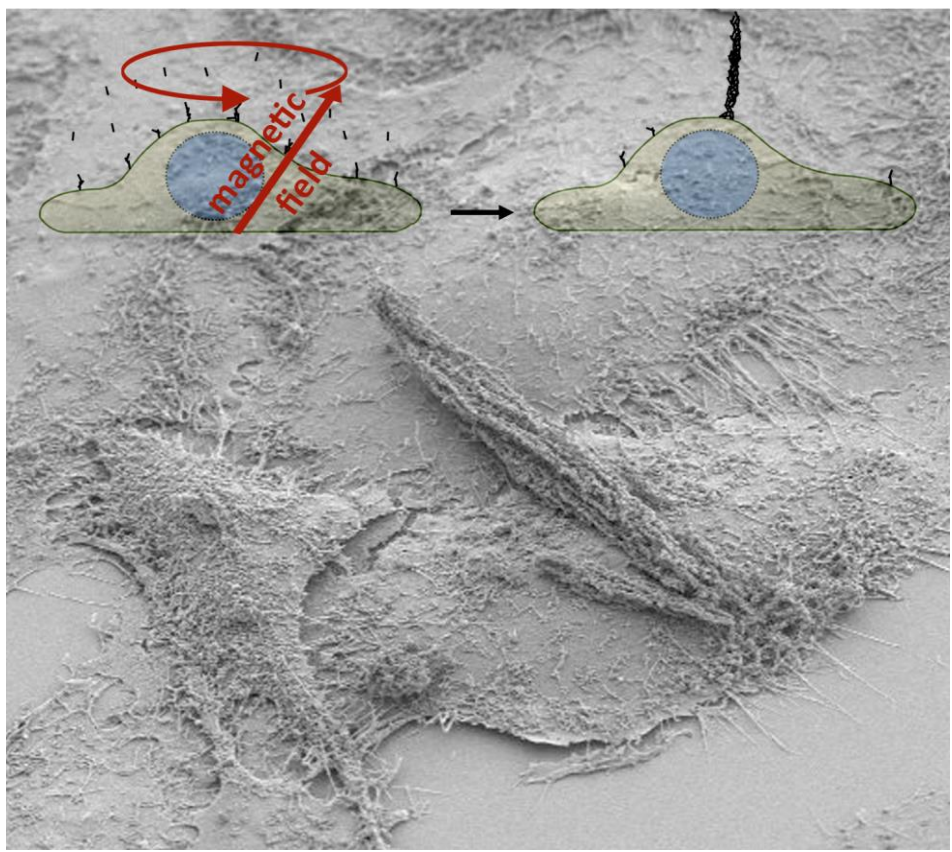